\documentclass[10pt,conference]{IEEEtran}

\usepackage{graphicx}
\usepackage{blkarray}
\usepackage{array}
\usepackage{multirow}
\usepackage{amsmath}
\usepackage{amsfonts}
\usepackage{amssymb}
\usepackage{amsthm}
\usepackage{arydshln}
\usepackage{color}

\newcommand{\Define}{\triangleq}

\begin{document}
\twocolumn

\title{Space-Time Index Modulation}
\author{Swaroop Jacob, T. Lakshmi Narasimhan$^{\dagger}$, and 
A. Chockalingam \\
Department of ECE, Indian Institute of Science, Bangalore 560012, India \\ 
$\dagger$ Presently with Department of EECS, Syracuse University, Syracuse, 
NY 13244, USA
}

\IEEEaftertitletext{\vspace{-0.6\baselineskip}}
\maketitle
\begin{abstract}
In this paper, we present a new multi-antenna modulation scheme, termed 
as {\em space-time index modulation (STIM)}. In STIM, information bits 
are conveyed through antenna indexing in the spatial domain, slot indexing 
in the time domain, and $M$-ary modulation symbols. A time slot in a given 
frame can be used or unused, and the choice of the slots used for 
transmission conveys slot index bits. In addition, antenna index bits are 
conveyed in every used time slot by activating one among the available 
antennas. $M$-ary symbols are sent on the active antenna in a used time 
slot. We study STIM in a cyclic-prefixed single-carrier (CPSC) system 
in frequency-selective fading channels. It is shown that, for the same 
spectral efficiency and single transmit RF chain, STIM can 
achieve better performance compared to conventional orthogonal frequency 
division multiplexing (OFDM). Low-complexity iterative algorithms for the 
detection of large-dimensional STIM signals are also presented. 
\end{abstract}

\section{Introduction}
\label{sec1}
Multi-antenna multiple-input multiple-output (MIMO) wireless systems 
are known to provide increased spectral and power efficiencies. 
Recently, various index modulation based schemes have been developed for 
MIMO systems to improve the achievable rate and performance with reduced 
hardware complexity \cite{sm1}-\cite{gsfim1}. Spatial modulation (SM) 
scheme performs index modulation in the spatial domain \cite{sm1}. SM 
activates only one transmit antenna among the available transmit antennas 
in any given channel use. Thus, SM reduces hardware complexity by using 
only one transmit radio frequency (RF) chain. The index of the active 
antenna used for transmission conveys information bits. SM was generalized 
to activate multiple transmit RF chains in a given channel use, which was 
termed as the generalized spatial modulation (GSM) \cite{gsm2a}, \cite{gsm3}. 
It has been shown that GSM outperforms conventional MIMO systems in terms of
bit error performance for a given spectral efficiency \cite{sm3}. 

Index modulation has been exploited in domains other than the spatial 
domain. Indexing subcarriers in multicarrier systems like OFDM, termed 
as subcarrier index modulation (SIM) \cite{sim1},\cite{sim2}, is an 
example of index modulation in the frequency domain. Generalized 
space-frequency index modulation (GSFIM) reported in \cite{gsfim1}, 
\cite{gsfm_det} combines the benefits of indexing over spatial and 
frequency domains. In GSFIM, information bits are conveyed through 
antenna indexing, subcarrier indexing, and modulation symbols.
However, GSFIM system does not employ antenna indexing in every channel 
use. The indices of the active antennas are fixed in a given GSFIM 
frame. This limits the rate achievable through antenna indexing. 
Generalized space-time shift keying (GSTSK) reported in \cite{gstsk}
performs indexing of space-time dispersion matrices. In GSTSK, the 
sum of a subset of all possible space-time dispersion matrices are 
chosen for transmission in a certain number of channel uses. The 
choice of this subset conveys information bits.

In this paper, we propose a new index modulation scheme, referred to 
as space-time index modulation (STIM). STIM efficiently performs 
indexing in spatial and time domains. In STIM, transmission is 
carried out in frames. Each frame consists of certain number of time 
slots. Not all the time slots in a frame are necessarily used for
transmission (i.e., no transmission takes place in certain time slots). 
In fact, the choice of the combination of used slots and unused 
slots in a frame conveys information bits through time slot indexing.
On the used time slots, modulation symbols are sent on the transmit
antenna chosen based on antenna index bits. Thus, information bits are 
conveyed through indices of the used time slots, index of
the active antenna, and modulation symbols. 

STIM is well suited for use in block transmission schemes. For example, 
STIM can be used in cyclic-prefixed single-carrier (CPSC) scheme 
\cite{cpsc2}, which is a block transmission scheme suited for inter-symbol 
interference (ISI) channels. In this context, the following two questions 
arise:
\begin{enumerate}
\item how does STIM used in a CPSC scheme compare in terms of 
rate and performance relative to conventional OFDM in ISI channels, and 
\item how to detect STIM signals, particularly when the dimensionality 
(i.e., frame size) is large. 
\end{enumerate}
While the first question is aimed at seeing if there are benefits 
vis-a-vis conventional schemes, the second question is aimed at 
addressing the implementation complexity issue. This paper presents 
answers to the above questions. First, it is found that, for the same 
spectral efficiency and single transmit RF chain, STIM in 
CPSC can outperform conventional OFDM. Second, low-complexity STIM detection 
algorithms which scale well for large dimensions are proposed. These 
results suggest that STIM can be a promising modulation scheme, and 
has the potential for further investigations beyond what is reported 
in this paper.

The rest of this paper is organized as follows. In Sec. \ref{sec2}, we 
present the STIM scheme, system model, and rate analysis. 
The proposed low-complexity algorithms for STIM signal detection are 
presented in Sec. \ref{sec3}. Bit error performance results and discussions 
are presented in Sec. \ref{sec4}. Conclusions and scope for 
future work are presented in Sec. \ref{sec5}.
\begin{figure*}[t]
\centering
\includegraphics[width=6.8in]{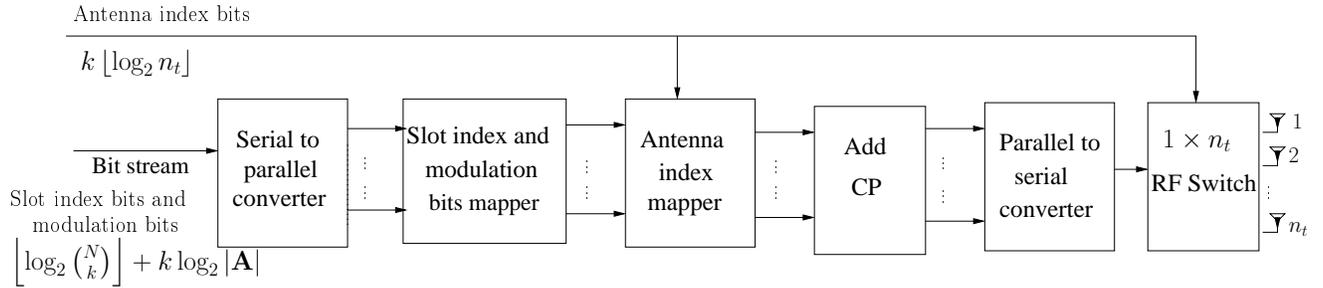}
\vspace{-2mm}
\caption{STIM transmitter.}
\vspace{-2mm}
\label{Fig_sys_model}
\end{figure*}

\section{STIM system model}
\label{sec2}	
The STIM scheme has $n_t$ transmit antennas with a single transmit RF 
chain (i.e., $n_{rf}=1$, where $n_{rf}$ is the number of transmit RF 
chains), and $n_r$ receive antennas. The channel is assumed 
to be frequency-selective with $L$ multipaths\footnote{The multipath 
delays can take generic delay values. As a result of Nyquist sampling 
at the receiver, the path delays can be reduced to integer multiples of 
the signaling interval \cite{tse}. The powers of the channel gains at 
each integer multiple of the signaling interval is given by the 
power-delay profile (PDP). The PDP of the frequency selective fading
channel captures the statistical characteristics and dependencies between
the various multipaths of the channel \cite{tse}. We consider an exponential
PDP as described in Sec. \ref{sec2c}.}. Transmission is carried 
out in frames. Each frame consists of $N+L-1$ channel uses, where $N$ 
denotes the length of the data part in number of channel uses and 
$L-1$ channel uses are used for transmitting cyclic prefix (CP). 
Information bits are conveyed through three different entities, namely, 
$i)$ indices of the used time slots in a frame, and $ii)$ index of the 
active transmit antenna in a channel use, and $iii)$ symbols from a 
modulation alphabet ${\mathbb A}$, as follows.

\subsection{Time slot indexing}	
\label{sec2a}
We have $N$ time slots in a frame available for conveying information. 
Among these $N$ slots, only $k$ slots are selected for transmission of 
modulation symbols. Let $\mathbb{A}$ denote the alphabet from which
the modulation symbols are drawn. There are ${N \choose k}$ possibilities 
of choosing which $k$ slots are used and which $N-k$ slots are not used. 
Specifically, a time slot is said to be used if any one of the $n_t$ 
antennas transmits a symbol from ${\mathbb A}$ in that slot. A slot is 
said to be not used, if none of the antennas transmit in that slot (i.e., 
all antennas remain silent, where silence can be viewed as sending a 0). 
The $k$ slots used for transmission in a given frame are chosen based on 
$\big \lfloor \log_2{N \choose k} \big\rfloor$ information bits.
Let us call a given realization of the used/unused status of $N$ slots 
in a frame as a `slot activation pattern (SAP)'. Out of
${N \choose k}$ possible SAPs, only 
$2^{\left \lfloor \log_2{N \choose k} \right\rfloor}$ SAPs 
are used for slot indexing to convey  
$\big\lfloor \log_2{N \choose k}\big\rfloor$ bits. 

{\em Example}: Let $N=8$, $k=7$. So, ${N \choose k}=8$ 
and  $\left \lfloor \log_2{8\choose 7} \right\rfloor=3$  slot 
index bits. The possible SAPs here are:
{\footnotesize
\begin{eqnarray*}
\hspace{0.5mm}
\Big\{
&\hspace{-4mm}
\begin{bmatrix}
0& 1& 1& 1& 
1& 1& 1& 1
\end{bmatrix}, 
\begin{bmatrix}
1& 0& 1& 1& 
1& 1& 1& 1
\end{bmatrix},& \nonumber \\
&\hspace{-4mm}
\begin{bmatrix}
1& 1& 0& 1& 
1& 1& 1& 1
\end{bmatrix},
\begin{bmatrix}
1& 1& 1& 0& 
1& 1& 1& 1
\end{bmatrix},& \nonumber \\
&\hspace{-4mm}
\begin{bmatrix}
1& 1& 1& 1& 
0& 1& 1& 1
\end{bmatrix},
\begin{bmatrix}
1& 1& 1& 1& 
1& 0& 1& 1
\end{bmatrix},& \nonumber \\
&\hspace{-4.5mm}
\begin{bmatrix}
1& 1& 1& 1& 
1& 1& 0& 1
\end{bmatrix},
\begin{bmatrix}
1& 1& 1& 1& 
1& 1& 1& 0
\end{bmatrix}\ &\hspace{-4mm}\Big\}, \nonumber
\end{eqnarray*}
}

\vspace{-4mm}
\hspace{-6.5mm}
where 1's correspond to the location of the used time slots and 0's
correspond to the location of the unused time slots. Each used slot
carries one symbol from ${\mathbb A}$, and nothing gets transmitted
in unused slots. 

\subsection{Antenna indexing}
\label{sec2b}
A transmit antenna is said to be active in a given channel use if it
transmits a non-zero modulation symbol in that channel use. Similarly,
a transmit antenna is said to be inactive if it does not transmit any
modulation symbol in that channel use. The transmit antenna activated 
in a given channel use is chosen based on 
$\left\lfloor \log_2 n_t \right\rfloor$ information bits. 
Note that in an STIM frame, in each of the $k$ used time slots the 
index of the active antenna can be different. In the remaining $N-k$ 
slots, all the $n_t$ transmit antennas are inactive. Since only $k$ 
out of $N$ time slots are used in an STIM frame, 
$k\left\lfloor \log_2 n_t \right\rfloor$ information bits are 
conveyed through antenna indices in one STIM frame.

With the addition of antenna indexing, the chosen SAP in a given frame 
gets mapped on to a {\em space-time activation matrix} ${\bf A}$ of 
dimension $n_t \times N$, whose entries are zeros and ones; 
${\bf A}_{i,j}=1$ means a  symbol from ${\mathbb A}$ is sent on the 
$i$th antenna in the $j$th slot, $i=1,2,\cdots,n_t$, and $j=1,2,\cdots,N$; 
${\bf A}_{i,j}=0$ means 0 is sent on the $i$th antenna in the $j$th slot 
(i.e., $j$th slot is unused). Note that only $k$ columns in ${\bf A}$ will 
have a non-zero entry and the remaining $N-k$ columns will have only zero 
entries.

{\em Example}: Let $n_t$ = 2, $N = 8$, $k = 7$. The possible 
`antenna activation patterns (AAP)' are given by 
$\big\{[ 0, 1]^T, [1, 0]^T\big\}$, where a 1 corresponds to the 
index of the active antenna, and a 0 corresponds to the index 
of the inactive antenna. Therefore, 
$\left\lfloor \log_2 n_t \right\rfloor=\left\lfloor \log_2 2 \right\rfloor=1$ 
bit can be conveyed through the index of the active antenna in each of 
the $k=7$ used slots.

Let ${\bf B}$ denote the $n_t\times N$ STIM signal matrix without CP.
The matrix ${\bf B}$ is formed by using symbols from ${\mathbb A}$ in 
those entries of the ${\bf A}$ matrix where ${\bf A}_{i,j}=1$.
Now, the STIM signal matrix with CP, denoted by ${\bf X}$ and of
size $n_t\times N+L-1$, is formed by appending $L-1$ CP symbols
to ${\bf B}$. The columns of ${\bf X}$ are transmitted in $N+L-1$ 
channel uses.

{\em Example}: Consider a system with $n_t=2, N=8, k=7, L=2$, and 4-QAM.
Let $[0 1 1 0 1 0 1 0 0 1 0 1 0 0 1 1 1 1 0 0 0 1 1 0]$ denote a 
possible input information bit sequence to be transmitted by the STIM 
transmitter. The first $k\left\lfloor \log_2 n_t\right\rfloor=7$ bits 
choose the index of the active antennas in the $k=7$ used time slots. 
The next $\big\lfloor \log_2{N \choose k}\big\rfloor=3$ bits choose 
the indices of the seven used time slots. The last $k\log_2|{\mathbb A}|=14$ 
bits choose the seven 4-QAM symbols. For this input bit sequence, the 
$\bf{A}$ matrix can be 
\[
{\bf A}= \begin{bmatrix}
1 & 0 & 0 & 1 & 0 & 1 & 0 & 0 \\
0 & 1 & 1 & 0 & 1 & 0 & 0 & 1 \end{bmatrix}, 
\] 
and the corresponding ${\bf B}$ matrix is 
\[ \scriptsize{{\bf B}}= \begin{bmatrix}
1- \mbox{j}  & 0 & 0 & -1-\mbox{j} & 0 & 1-\mbox{j} & 0 & 0\\
0 & 1+\mbox{j} & -1-\mbox{j} & 0 & 1+\mbox{j} & 0 & 0 & -1+\mbox{j} \end{bmatrix},
\]
where $\mbox{j}=\sqrt{-1}$. The STIM signal matrix with CP appended is then
given by 
\[\scriptsize{{\bf X}\hspace{-0mm}=\hspace{-0mm}\begin{bmatrix}
0&\hspace{-1mm}1-\mbox{j}  &\hspace{-1mm} 0 &\hspace{-1mm} 0 &\hspace{-1mm} -1-\mbox{j} &\hspace{-1mm} 0 &\hspace{-1mm} 1-\mbox{j} &\hspace{-1mm} 0 &\hspace{-1mm} 0\\
-1+\mbox{j}&\hspace{-1mm}0 &\hspace{-1mm} 1+\mbox{j} &\hspace{-1mm} -1-\mbox{j} &\hspace{-1mm} 0 &\hspace{-1mm} 1+\mbox{j} &\hspace{-1mm} 0 &\hspace{-1mm} 0 &\hspace{-1mm} -1+\mbox{j} \end{bmatrix}}.
\]
The block diagram of STIM transmitter is shown in Fig. \ref{Fig_sys_model}.

\subsection{Analysis of achieved rate in STIM}
In this subsection, we analyze the achieved rate in STIM. We also compare
the rates achieved in STIM and conventional OFDM, both using a single 
transmit RF chain. Based on the description of STIM in the previous
subsection, the achieved rate in STIM can be written as 
\begin{eqnarray}
R_{\mbox{{\scriptsize STIM}}}&\hspace{-1mm} = & \hspace{-1mm}
\frac{1}{N+L-1} \bigg[ \underbrace{k{\left\lfloor \log_2{n_t}\right\rfloor}}_{\mbox{antenna index bits}}
+ \underbrace{\Big\lfloor \log_2{N \choose k}\Big\rfloor }_{\mbox{slot index bits}} \nonumber \\
& & + \ \underbrace{k\log_2 |{\mathbb A}|}_{\mbox{modulation symbol bits}} \bigg] \quad \mbox{bpcu}.
\label{stim_rate1}
\end{eqnarray}
Note that in conventional OFDM, there are no slot and antenna index bits 
to contribute to the achieved rate. All the subcarriers carry modulation 
symbols. The achieved rate in OFDM with $N$ subcarriers is given by
\begin{equation}
R_{\mbox{{\tiny OFDM}}} \ = \ \frac{N\log_2{|{\mathbb A}|}}{N+L-1} \ \ \mbox{ bpcu}.
\label{ofdmrate}
\end{equation}
From (\ref{stim_rate1}) and (\ref{ofdmrate}), we can write the percentage 
of rate improvement offered by STIM over OFDM as 
\begin{equation}
\mbox{Rate improvement } (R_I) = 
\frac{R_{\mbox{{\tiny STIM}}}- R_{\mbox{{\tiny OFDM}}}}{R_{\mbox{{\tiny OFDM}}}}
\times 100 \ \%.
\label{ri}
\end{equation}

\subsubsection{Maximizing $R_{STIM}$ over $k$}
Define $M\Define\log_2|{\mathbb A}|$ and $A\Define\left\lfloor\log_2{n_t}\right\rfloor$.
From (\ref{stim_rate1}), the achieved rate in STIM as a function 
of $k$ can be written as
\begin{equation}
R(k)=\frac{k(A+M)+\log_2{N \choose k}}{N+L-1}.
\label{rk}
\end{equation}
We need to obtain a $k$ that maximizes $R(k)$. The combinatorial nature of the
expression for $R(k)$ makes it difficult to obtain the value of $k$ that
maximizes $R(k)$ through differentiation technique. Thus, we try to obtain
a range of values for $k$ over which $R(k)$ is maximized.

Let $k^*$ be a value of $k$ for which $R(k)$ is maximum. Since $R(k)$ is
a concave function\footnote{It is known that ${N \choose k}$ 
is concave w.r.t $k$ and $\log(.)$ is a monotone function. Therefore, from 
(\ref{rk}), $R(k)$ is a concave function w.r.t $k$.}, $k^*$ should satisfy
\begin{eqnarray}
R(k^*)-R(k^*-1)& \hspace{-2mm} \geq& \hspace{-2mm} 0, \label{rc1}\\
R(k^*+1)-R(k^*)&\hspace{-2mm} \leq& \hspace{-2mm} 0.
\label{rc2}
\end{eqnarray}
Expanding (\ref{rc1}), we write
\begin{eqnarray}
R(k^*)-R(k^*-1)&\hspace{-2mm} \geq& \hspace{-2mm} 0 \nonumber\\
\Rightarrow
k^*(A+M)+\log_2{N \choose k^*} \nonumber \\
-(k^*-1)(A+M)
-\log_2{N \choose k^*-1}&\hspace{-2mm} \geq& \hspace{-2mm}0 \nonumber\\
\Rightarrow
A+M & \hspace{-2mm} \geq& \hspace{-2mm} \log_2\frac{{N \choose k^*-1}}{{N \choose k^*}}\nonumber\\
\Rightarrow
A+M & \hspace{-2mm} \geq& \hspace{-2mm} \log_2\frac{k^*}{N-k^*+1}\nonumber\\
\Rightarrow
2^{A+M} & \hspace{-2mm}\geq& \hspace{-2mm} \frac{k^*}{N-k^*+1}. \label{ub1}
\end{eqnarray}
Define $C\Define 2^{A+M}$. From (\ref{ub1}), an upper bound for $k^*$ can
be written as
\begin{equation}
k^*\leq \ k_u \ \Define \ \frac{C(N+1)}{1+C}.
\end{equation}
Next, we obtain a lower bound by expanding (\ref{rc2}) as follows:
\begin{eqnarray}
R(k^*+1)-R(k^*)& \hspace{-2mm} \leq& \hspace{-2mm} 0 \nonumber\\
\Rightarrow
(k^*+1)(A+M)+\log_2{N \choose k^*+1} \nonumber \\
-(k^*)(A+M)-\log_2{N \choose k^*}& \hspace{-2mm}\leq& \hspace{-2mm}0 \nonumber\\
\Rightarrow
A+M & \hspace{-2mm}\leq& \hspace{-2mm}\log_2\frac{{N \choose k^*}}{{N \choose k^*+1}}\nonumber\\
\Rightarrow
A+M & \hspace{-2mm} \leq& \hspace{-2mm} \log_2\frac{k^*+1}{N-k^*}\nonumber\\
\Rightarrow
C & \hspace{-2mm}\leq& \hspace{-2mm} \frac{k^*+1}{N-k^*}.
\end{eqnarray}
Now, a lower bound for $k^*$ can be written as
\begin{equation}
k^* \ \geq \ k_l \ \Define \ \frac{CN-1}{1+C}.
\label{lb1}
\end{equation}
From (\ref{ub1}) and (\ref{lb1}), we can see that any $k$ such that
$k_l\leq k \leq k_u$ will maximize $R(k)$. The bounds $k_l$ and
$k_u$ are considerably tight. This can be seen by
\begin{equation}
k_u-k_l \ = \ \frac{C(N+1)-CN+1}{1+C} \ = \ 1.
\end{equation}
Therefore, maximum rate can be achieved in STIM when we choose the
value of $k$ to be $k_m$, where
\begin{equation}
k_m \ = \ \frac{k_l+k_u}{2} \ = \ \frac{C(N+1)}{1+C}-\frac{1}{2}.
\label{km}
\end{equation}

\subsubsection{Maximizing $R_I$ over $N$}
From (\ref{stim_rate1}), (\ref{ofdmrate}), and (\ref{ri}), the percentage 
rate improvement $R_I$ can be written as
\begin{eqnarray}
R_I &=&\frac{k(A+M)+\log_2{N \choose k}-NM}{NM}.
\label{ri1}
\end{eqnarray}
Let us assume\footnote{
This assumption can be reasoned as follows: for large values of $C$, from
(\ref{lb1}), the value of $k^*$ can be approximated to the nearest integer 
as $k^* \approx N-1$. Further, this can be observed in Fig. \ref{maxrs}, 
where, as $C$ (or alternatively $M$) increases, the value of $k$ for which 
the maximum rate $R_{\scriptsize{\mbox{STIM}}}$ is achieved increases and 
tends towards $N-1$.} $k=N-1$.  Now, (\ref{ri1}) can be written as
\begin{eqnarray}
R_I &=&\frac{(N-1)(A+M)-NM+\log_2N}{NM}.\nonumber
\end{eqnarray}
To maximize $R_I$ with respect to $N$, we evaluate $\frac{dR_I}{dN}=0$:
\begin{eqnarray}
\frac{dR_I}{dN} & \hspace{-1mm} =& \hspace{-1mm} \frac{A+M}{N^2M}-\frac{\log_2N}{N^2M}+\frac{1}{N^2M\log(2)}\ = \ 0,\nonumber\\
\Rightarrow  \hspace{2mm} \log_2N& \hspace{-1mm} =& \hspace{-1mm} A+M+1.4427,\nonumber\\
\Rightarrow \hspace{2mm}N& \hspace{-1mm} =& \hspace{-1mm} C2^{1.4427} \ = \ n_t|{\mathbb A}|2.7183.
\label{ri2}
\end{eqnarray}
Therefore, the percentage of rate improvement obtained through STIM over
OFDM can be maximized when the number of slots $N=C2^{1.4427}$.

\subsubsection{Numerical results}
\begin{figure}
\centering
\includegraphics[width=3.75in, height=2.75in]{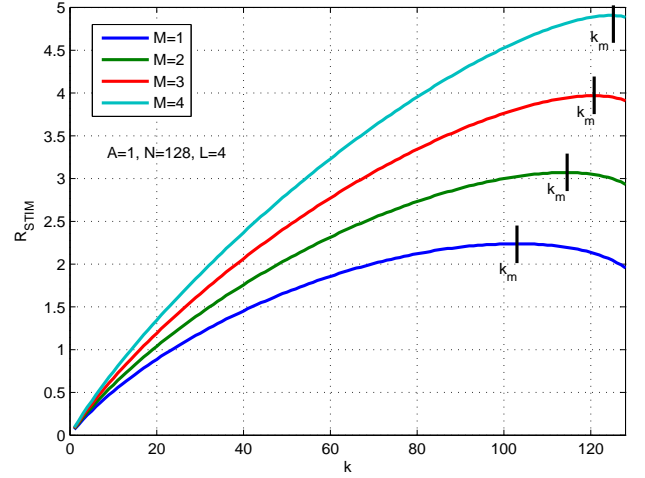}
\vspace{-6mm}
\caption{Achieved rate of STIM as a function of $k$
for $N=128$, $n_t=2$, $n_{rf}=1$, and $|{\mathbb A}|=2,4,8,16$.}
\label{maxrs}
\vspace{-4mm}
\end{figure}

In Fig. \ref{maxrs}, we plot the variation of the achieved rate of STIM 
as a function of $k$. This figure shows the values of $R(k)$ for varying 
$k\in\{1,2,\cdots, N\}$, $N=128$, $n_t=2$ (i.e., $A=1$), $n_{rf}=1$, $L=4$, 
and $|{\mathbb A}|=2,4,8,16$ (i.e., $M=1,2,3,4$). It can be seen that the 
STIM rate is maximized only by certain values of $k$. Further, the values 
of $k_m$ given by (\ref{km}) coincide with the values of $k$ that maximize 
$R(k)$. For $M=1,2,3,4$, the values of $k_m$ are $103, 114, 121, 125$, 
respectively.

\begin{figure}
\centering
\includegraphics[width=3.75in, height=2.75in]{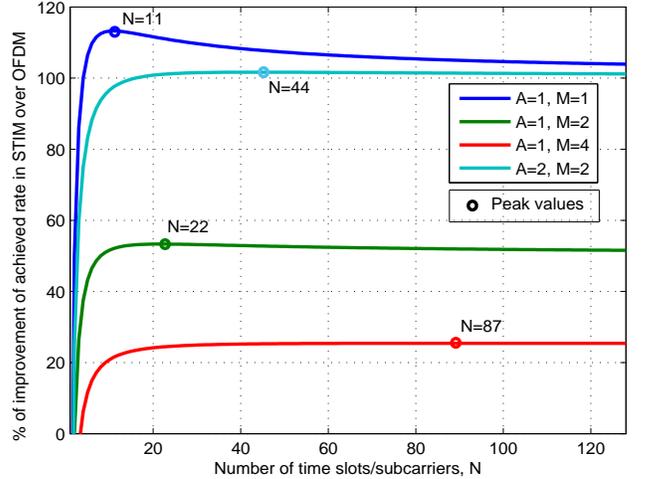}
\vspace{-6mm}
\caption{Percentage rate improvement in STIM over OFDM for 
varying number of slots $N$, 
and varying $n_t$ and $\mathbb A$.}
\label{maxri}
\vspace{-4mm}
\end{figure}

In Fig. \ref{maxri}, we plot $R_I$ as a function of $N$. This figure 
presents the values of $R_I$ for 
$i$) $n_t=2$ (i.e., $A=1$), 
and $|{\mathbb A}|=2$ (i.e., $M=1$), $ii$) $n_t=2$ ($A=1$), and 
$|{\mathbb A}|=4$ ($M=2$), $iii$) $n_t=2$ ($A=1$), and $|{\mathbb A}|=16$ 
($M=4$), and $iv$) $n_t=4$ ($A=2$), and $|{\mathbb A}|=4$ ($M=2$). The 
value of $N$ obtained through simulation, that maximizes $R_I$, matches 
well with the analytical expression given by (\ref{ri2}). For the system 
parameters considered in Fig. \ref{maxri}, $R_I$ is maximum at
$i$) $N=2^{3.4427}\approx 11$,
$ii$) $N=2^{4.4427}\approx 22$,
$iii$) $N=2^{6.4427}\approx 87$, and
$iv$) $N=2^{5.4427}\approx 44$.

\subsection{Received STIM signal}
\label{sec2c}
Now, let us write the received STIM signal model at the receiver.
Let $h^{(j,i)}(l)$ denote the channel gain from $i$th transmit antenna to 
the $j$th receive antenna on the $l$th multipath. The power delay profile 
of the channel is assumed to follow exponential decaying model, i.e., 
$\mathbb{E}[|h^{(j,i)}(l)|^2]=e^{-l}$, $l=0,1,\cdots,L-1$. 
Let $\mathbf{x}_i$ denote the transmitted symbol vector of size $n_t\times 1$
in the $i$th channel use, $1\leq i\leq N+L-1$. Note that $\mathbf{x}_i$ is 
the $i$th column of the matrix ${\bf X}$. We assume that the channel remains 
invariant in one STIM frame duration. At the receiver, after removing the 
cyclic prefix, the received vector can be represented as 
\begin{eqnarray}
\mathbf{y} &=& \mathbf{H} \mathbf{x}+\mathbf{n},
\label{sys_new1}
\end{eqnarray}
where $\mathbf{n}$ is the noise vector of size $Nn_r \times 1$ with 
$\mathbf{n}\sim {\cal{CN}}(0,\sigma^2\mathbf{I}_{Nn_r})$, $\mathbf{x}$ is 
the vector of size $Nn_t \times 1$ given by
$\mathbf{x} = \begin{bmatrix}\mathbf{x}_L^T & \mathbf{x}_{L+1}^T & \ldots & \mathbf{x}_{N+L-1}^T\end{bmatrix}^T $, 
and $\mathbf{H}$ is the $Nn_r \times Nn_t$ equivalent block circulant 
channel matrix given by
\[{\scriptsize
\mathbf{H}\hspace{-0.5mm} = \hspace{-0.5mm} \begin{bmatrix} \mathbf{H}_{0} & 0 & 0 & 0 & \cdots & 0 & \mathbf{H}_{L-1} & \cdots  & \mathbf{H}_{1}\\
\mathbf{H}_{1} & \mathbf{H}_{0} & 0 & 0 & \cdots & 0 & 0 & \cdots &  \mathbf{H}_{2}\\
\vdots & \vdots & \vdots & \vdots & \vdots & \vdots & \vdots  & \vdots\\
\mathbf{H}_{L-2} & \mathbf{H}_{L-3} & \mathbf{H}_{L-4} & \cdot & \mathbf{H}_{0} & 0 & \cdot & \cdots &  \mathbf{H}_{L-1}\\
\mathbf{H}_{L-1} & \mathbf{H}_{L-2} & \mathbf{H}_{L-3} & \cdot & \mathbf{H}_{1} & \mathbf{H}_{0} & 0 & \cdots &  0\\
0 & \mathbf{H}_{L-1} & \mathbf{H}_{L-2} & \cdot & \mathbf{H}_{2} & \mathbf{H}_{1} & \mathbf{H}_{0} & \cdots &  0\\
\mbox{\ensuremath{\vdots}} & \vdots & \vdots & \vdots & \vdots & \vdots & \vdots & \vdots\\
0 & 0 & 0 & \cdot & \cdot & \cdot & \cdot & \cdots &  \mathbf{H}_{0},
\end{bmatrix}},
\]
where $\mathbf{H}_{l}$ is the $n_r \times n_t$ channel matrix corresponding
to the $l$th multipath. 
Assuming perfect knowledge of ${\bf H}$ at the 
receiver, the maximum likelihood (ML) detection rule is given by 
\begin{align}
\hat{\mathbf{x}} = \underset{\mathbf{x} \in \mathbb{S}} {\text{argmin}} \ \|\mathbf{y}-\mathbf{H}\mathbf{x}\|^2,
\label{ml_eq}
\end{align}
where ${\mathbb S}$ denotes the set of all possible ${\bf x}$ vectors.

\section{Low-complexity STIM detection}
\label{sec3}
From (\ref{ml_eq}), it can be seen that the ML detection of an STIM 
frame has a complexity that increases exponentially with $N$, $n_t$. 
In order to enable the detection of large dimensional 
STIM signals, in this section, we present two low-complexity detection 
algorithms using message passing. 

\subsection{Two-stage STIM detector}
\label{sec3a}
In this proposed two-stage STIM detector (2SSD), detection is carried out 
in two stages. The first stage is a minimum mean square error (MMSE) 
estimator, which is followed by a message passing based detector. The 
purpose of the first stage detection is to provide an estimate of the 
indices of the active antennas in the STIM frame. Using this estimate, 
the message passing based detector obtains an estimate of the indices
of the used time slots in the frame and the modulation symbols. Finally, 
these estimates are demapped to obtain an estimate of the transmitted 
information bit sequence.
	
{\em Stage 1} : The first stage detection is performed using MMSE
estimator as 
\begin{equation}
\hat{\mathbf{x}} \ = \ \left[ \mathbf{H}^H\mathbf{H}+\sigma^2I\right]^{-1}\mathbf{H}^H\mathbf{y},
\end{equation}
where $\hat{\mathbf{x}}$ is of the form
$\hat{\mathbf{x}}=\begin{bmatrix}\hat{\mathbf{x}}_L^T & \hat{\mathbf{x}}_{L+1}^T & \ldots & \hat{\mathbf{x}}_{N+L-1}^T\end{bmatrix}^T$.
We construct the index set ${\mathcal I}=\{I_L,I_{L+1},\cdots,I_{N+L-1}\}$, 
where $I_i$ is the index of the element with the largest magnitude in 
$\hat{\mathbf{x}}_i$, and $L\leq i\leq N+L-1$. 
The set ${\mathcal I}$ gives the estimates of the indices of the active 
antennas in the $N$ channel uses of an STIM frame. Now, we can write
(\ref{sys_new1}) in the form 
\begin{eqnarray}
\mathbf{y}&=&\mathbf{\bar H} \mathbf{z}+\mathbf{n},
\label{sys_new2}
\end{eqnarray}
where $\bf z$ is an $N \times 1$ vector whose elements correspond to the 
elements in $\bf x$ at the indices given by the set ${\mathcal I}$, and 
$\mathbf{\bar H}$ is the $Nn_r \times N$ channel matrix obtained by 
choosing the columns of $\mathbf{H}$ at the indices given by the set 
${\mathcal I}$.

{\em Stage 2}: The second stage of the 2SSD algorithm is a message 
passing based algorithm that works on the model given by (\ref{sys_new2}). 
The message passing algorithm gives the estimates of the indices of
the used time slots and the modulation symbols. 

Let $t_i$ be the time slot activity indicator variable for the $i$th 
time slot. That is, $t_i=1$ whenever $i$th time slot is used, and $t_i=0$ 
otherwise. The $i$th time slot corresponds to the $i$th entry of vector 
$\mathbf{z}$. Note that $\sum_{i=1}^{N}t_i=k$. This is referred to as the 
STIM time constraint $\mathbf{T}$. Let 
$\mathbf{t}\triangleq[t_1,t_2,\cdots,t_N]$. We obtain the a posteriori 
probabilities (APP) of the elements of $\mathbf{t}$ and $\mathbf{z}$, 
using which the time slot index bits and the modulation symbol bits are 
estimated.

\begin{figure}[t]
\centering
\includegraphics[width=2.8in, height=1.75in]{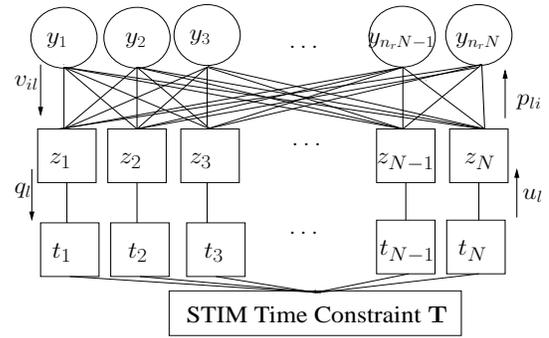}
\caption{The graphical model and messages in the second stage 
messaging passing algorithm of 2SSD.}
\label{Fig_graph_model}
\vspace{-2mm}
\end{figure}
	
Figure \ref{Fig_graph_model} shows the graphical model for message passing.
The graph consists of four sets of nodes. Messages are exchanged between 
these nodes in two layers: one corresponding to the probabilities of the 
modulation symbols and the other corresponding to the probabilities of 
the time slot activity indicators. For the system described in 
(\ref{sys_new2}), the a posteriori probability  is given by
\begin{eqnarray}    
\mbox{Pr}(\mathbf{z}|\mathbf{y}) &\hspace{-0.0mm}=& \hspace{-0.0mm}\mbox{Pr}(\mathbf{z},\mathbf{t}|\mathbf{y}) \nonumber \\
& \hspace{-0.0mm}\propto & \hspace{-0.0mm} \mbox{Pr}(\mathbf{y}|\mathbf{z},\mathbf{t}) \mbox{Pr}(\mathbf{z},\mathbf{t}) \nonumber \\
& = & \mbox{Pr}(\mathbf{y}|\mathbf{z})\mbox{Pr}(\mathbf{z}|\mathbf{t})\mbox{Pr}(\mathbf{t}) \nonumber\\
&\hspace{-0.0mm}=&\hspace{-0.0mm}\bigg \{\prod_{j=1}^{n_rN} \mbox{Pr}(y_j|\mathbf{z})\prod_{i=1}^{N}\mbox{Pr}(z_i|t_i)\bigg \}\mbox{Pr}(\mathbf{t}).
\label{eqn:prob1}
\end{eqnarray}
The graph is constructed so as to marginalize the above probability 
distribution. The following four sets of nodes are defined:
$i)$ $n_rN$ observation nodes corresponding to the elements of $\mathbf{y}$, 
$ii)$ $N$ variable nodes corresponding to the elements of $\mathbf{z}$, 
$iii)$) $N$ time slot activity nodes corresponding to the elements of 
$\mathbf{t}$, and $iv)$ a constraint node corresponding to the time
constraint $\mathbf{T}$.
The messages that are exchanged in layer 1 between the observation and 
the variable nodes produce approximate APPs of the individual elements 
of $\mathbf{z}$. The messages that are exchanged in layer 2 between the 
time slot activity nodes and the STIM time constraint node generate the 
APPs of the elements of $\mathbf{t}$. To generate these messages, we 
employ a Gaussian approximation of interference as follows.
We write the $i$th element of ${\bf y}$ as
\begin{eqnarray}
y_i&=&{\bar H}_{i,l}{z}_l +
\underbrace{\sum_{j=1,j\neq l}^{N}{\bar H}_{i,j}{z}_j+{n}_i}_{\triangleq\tiny{d_{i,l}}},
\end{eqnarray}	
where $i=1,2,\cdots,n_rN$ and $l=1,2,\cdots,N$. We approximate $d_{i,l}$ 
to be Gaussian with mean $\mu_{i,l}$ and variance $\sigma^2_{i,l}$,
which are calculated as
\begin{equation}
\mu_{i,l} \ = \ \mathbb{E}\Bigg [\sum_{{j=1,}\atop{j\neq l}}^{N}{\bar H}_{i,j}{z}_j+{n}_i\Bigg ], 
\end{equation}
and
\begin{equation}
\sigma^2_{i,l} \ = \ \mbox{Var}\bigg(\hspace{-0.5mm}\sum_{{j=1,}\atop{j\neq l}}^{N}{\bar H}_{i,j}{z}_j\hspace{-0.5mm}\bigg) +\sigma^2.
\end{equation}
The messages passed between the nodes are given below.

{\em Layer 1:} The message $v_{il}$ computed at the observation node is 
\begin{eqnarray}
v_{il}(z) & \triangleq & \text{Pr}(z_l=z|y_i)\nonumber \\
& \approx & \frac{1}{\sigma_{i,l}\sqrt{2\pi}}\mbox{exp}\bigg (\frac{-(y_i-\mu_{i,l}-z{\bar H}_{i,l})^2}{2\sigma^2_{i,l}}\bigg),
\end{eqnarray}
where $z\in{\mathbb A}\cup 0$. The APP of the individual elements of 
$\mathbf{z}$ is obtained at the variable nodes as
\begin{eqnarray}
\hspace{-5mm}
p_{li}(z) &\triangleq& \text{Pr}(z_l=z|\mathbf{y}_{\setminus i}) \nonumber\\
&\approx& \hspace{-2mm}\prod_{j=1,j\neq i}^{n_rN}\text{Pr}(z_l=z|y_j) 
\propto u_l(z^{\odot})\hspace{-1mm}\prod_{j=1,j\neq i}^{n_rN}\hspace{-1mm}v_{jl}(z),
\end{eqnarray}
where $\mathbf{y}_{\setminus i}$ denotes the set of all elements of 
$\mathbf{y}$ except $y_i$, and $z^{\odot}=0$ if $z=0$, and 
$z^{\odot}=1$ if $z\neq 0$. 
	
{\em Layer 2:} The APP estimate of $t_l$ from the variable nodes are 
computed as
\begin{eqnarray}
q_l(b)&\triangleq&\mbox{Pr}(t_l=b|{\bf z})\nonumber\\
&\propto& \begin{cases}\sum_{z \in \mathbb{A}}\prod_{j=1}^{n_rN}v_{jl}(z), &\mbox{if} \hspace{2mm} b=1\\ \prod_{j=1}^{n_rN}v_{jl}(0)&\mbox{if}\hspace{2mm}b=0.
\end{cases}
\end{eqnarray}
The APP estimate of $t_l$ computed at the time slot activity nodes after 
processing the STIM time constraint $\mathbf{T}$ is
\begin{eqnarray}
u_l(b)&=& \mbox{Pr}(t_l=b|\mathbf{z}_{\setminus l})\nonumber\\
&\propto& \begin{cases} \mbox{Pr}(\sum_{j=1,j\neq l}^{N}t_j=k-1|\mathbf{t}_{\setminus l}), &\mbox{if}\hspace{2mm} b=1 \\ \mbox{Pr}(\sum_{j=1,j\neq l}^{N}t_j=k|\mathbf{t}_{\setminus l}), &\mbox{if} \hspace{2mm}b=0 
\end{cases}\nonumber \\
&\approx& \begin{cases} \phi_{l}(k-1) &\mbox{if}\hspace{2mm} b=1 \\ \phi_{l}(k) &\mbox{if} \hspace{2mm}b=0,
\end{cases}
\end{eqnarray}
where $\mbox{Pr}(\sum_{j=1,j\neq {l}}^{N}t_j=k-1|\mathbf{t}_{\setminus l})$ 
denotes the probability that the time slot activity pattern satisfies 
the STIM time constraint $\mathbf{T}$ given that the $l$th time slot 
is used, and 
$\mbox{Pr}(\sum_{j=1,j\neq {l}}^{N}t_j=k|\mathbf{t}_{\setminus i})$ 
denotes the probability that the time slot activity pattern satisfies 
$\mathbf{T}$ given that the $l$th time slot is unused. This probability 
is evaluated as $\phi_{l}=\otimes_{j=1,j\neq l}^{N}q_j$, where $\otimes$ 
is the convolution operator, and $\phi_l(.)$ is a probability mass 
function with probability masses at $N$ points $(0,1,\cdots,N-1)$.
The messaging passing algorithm is summarized as follows.
\begin{itemize}
\item[1.] Initialize $p_{li}(z)=\frac{1}{|\mathbb{A}_0|}$, 
$q_l(1)=\frac{k}{N}$, $q_l(0)=1-\frac{k}{N}$, $\forall z,i,l$.
\item[2.] Compute $v_{il}(z)$, $\forall z,l,i$.
\item[3.] Compute $u_l(b)$, $\forall l,b$.
\item[4.] Compute $p_{li}(z)$, $\forall l,i,z$.
\item[5.] Compute $q_l(b)$, $\forall l,b$.
\end{itemize}
The above steps are repeated for a fixed number of iterations. 
Damping of the messages $p_{li}$'s and $q_{l}$'s is done using 
a damping factor $\Delta \in (0,1]$ in each iteration \cite{damp}. 
At the end of the iterations, an estimate of the index of the used 
time slots are obtained by choosing the $k$ time slots that have the 
largest APP. That is, a set ${\mathcal T}=\{T_1,T_2,\cdots,T_k\}$ 
is obtained such that $\{q_{T_1}(1),q_{T_2}(1),\cdots,q_{T_k}(1),\}$ are 
the $k$ largest APP values of $q_l(1),\ 1\leq l\leq N$. These estimates 
are demapped to get the time slot index bits. Since the transmit antennas 
are activated only in $k$ time slots, the $k$ antenna indices from the set 
$\mathcal I$ that correspond to the $k$ used time slot estimates are 
chosen and demapped to obtain the antenna index bits. Finally, from 
$p_{li}$'s, the information bits corresponding to the $k$ modulation 
symbols are obtained.
	
\subsection{Three-stage STIM detector}
\label{sec3b}
In this subsection, we present another detection algorithm called 
three-stage STIM detection (3SSD) algorithm, where the first two stages 
are the same as those in 2SSD. The third stage in the 3SSD aims to further
improve the reliability of the estimated antenna index bits and the 
modulation symbol bits through the use of message passing over a
bipartite (factor) graph. The estimates of the indices of the 
used time slots (i.e., the set $\mathcal T$) obtained from the second 
stage are fed as input to the third stage. The third stage achieves 
improved performance compared to 2SSD without much increase in complexity.  
The message passing scheme for the third stage is described below.
	
\begin{figure}[t]
\includegraphics[width=3.5in]{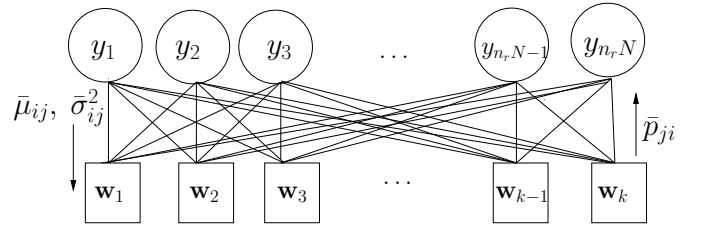}
\caption{The graphical model with the different messages passed in the 
third stage message passing algorithm of 3SSD.}
\label{Fig_graph_model2}
\end{figure}

Let $\mathbb M$ denote the set of all possible $n_t\times 1$ vectors 
that could be transmitted by the STIM transmitter in an used time 
slot. Therefore, $|{\mathbb M}|=n_t|{\mathbb A}|$. For example, for 
$n_t=2$ and BPSK, the set $\mathbb M$ is given by
\[
\mathbb{M} =
\left\lbrace
\begin{bmatrix}
1\\0
\end{bmatrix},
\begin{bmatrix}
-1\\0
\end{bmatrix},
\begin{bmatrix}
0\\1
\end{bmatrix},
\begin{bmatrix}
0\\-1
\end{bmatrix}
\right\rbrace.
\]
Let $\mathbf{w}_{j}\in {\mathbb M}$ be the $n_t$-length vector transmitted 
in the $j$th used time slot, and 
$\mathbf{w} = \begin{bmatrix}\mathbf{w}_1^T & \mathbf{w}_2^T & \cdots & \mathbf{w}_j^T & \cdots & \mathbf{w}_k^T \end{bmatrix}^T$.
Let $\mathbf{G}$ be an $n_rN\times kn_t$ matrix obtained by choosing 
$kn_t$ columns of the channel matrix $\mathbf{H}$ corresponding to 
the estimated indices of the $k$ used time slots given by the set 
$\mathcal T$. Now, the received signal can be expressed as
\begin{eqnarray}
\mathbf{y}&=&\mathbf{G} \mathbf{w}+\mathbf{n}.
\label{enh_1}
\end{eqnarray}
The third stage message passing works on the system model in (\ref{enh_1}), 
and estimates $\mathbf{w}$, given $\mathbf{y}$ and $\mathbf{G}$. The 
graphical model of this scheme consists of $k$ variable nodes each 
corresponding to a $\mathbf{w}_{j}$, and $Nn_r$ observation nodes each 
corresponding to a ${y_{i}}$. This graphical model is illustrated in Fig. 
\ref{Fig_graph_model2}. 
	
The messages passed between the variable nodes and the observation nodes 
are constructed as follows. From (\ref{enh_1}), the received signal 
$y_i$ can be written as
\begin{eqnarray}
y_i \hspace{-2mm}&=&\hspace{-2mm} \mathbf{g}_{i,[j]}{\mathbf{w}}_j +
\underbrace{\sum_{l=1,l\neq j}^{k}{\mathbf{g}}_{i,[l]}{\mathbf{w}}_l+{n}_i}_{\triangleq\tiny{f_{i,j}}},
\end{eqnarray}
where ${\mathbf{g}}_{i,[l]}$ is a row vector of length $n_t$ given by 
$\begin{bmatrix}G_{i,(l-1)n_t+1} & G_{i,(l-1)n_t+2} & \cdots & G_{i,ln_t}
\end{bmatrix}$,
and $\mathbf{w}_l \in \mathbb{M}$ is the $n_t$-length vector transmitted
in the $l$th used time slot. We approximate $f_{i,j}$ to be Gaussian 
with mean $\bar{\mu}_{i,j}$ and variance $\bar{\sigma}^2_{i,j}$, where	
\begin{eqnarray}
\bar{\mu}_{i,j}  \ =  \ \mathbb{E}\bigg [\sum_{{l=1,}\atop{l\neq j}}^{k}{\mathbf{g}}_{i,[l]}{\mathbf{w}}_l+{n}_i\bigg ] \ = \ \sum_{{l=1,}\atop{l\neq j}}^{k}\sum_{\mathbf{s}\in \mathbb{M}}\bar{p}_{li}(\mathbf{s}){\mathbf{g}}_{i,[l]}\mathbf{s},
\label{enh:mean1}
\end{eqnarray}
and
\begin{eqnarray}
\bar{\sigma}^2_{i,j}\hspace{-2mm}&\hspace{-2mm}=&\mbox{Var}\bigg(\sum_{{l=1,}\atop{l\neq j}}^{k}{\mathbf{g}_{i,[l]}{\mathbf{w}}_l+{n}_i\bigg )} \nonumber \\
&=&\sum_{{l=1,}\atop{l\neq j}}^{k}\bigg(\sum_{\mathbf{s}\in \mathbb{M}}\bar{p}_{li}(\mathbf{s})\mathbf{g}_{i,[l]}\mathbf{s}\mathbf{s}^H\mathbf{g}_{i,[l]}^H \nonumber \\
& & -\Big |\hspace{-0.5mm}\sum_{\mathbf{s}\in \mathbb{M}}\bar{p}_{li}(\mathbf{s})\mathbf{g}_{i,[l]}\mathbf{s}\Big |^2\bigg)+\sigma^2,
\label{enh:var1}
\end{eqnarray}
where $\bar{p}_{ji}(\mathbf{s})$ denotes the a posteriori probability 
message computed at the variable nodes as
\begin{align}
\bar{p}_{ji}(\mathbf{s}) & \propto \prod_{m=1,\ m\neq i}^{Nn_r} \mbox{exp}\bigg (\frac{-\left| y_m-\bar{\mu}_{m,j}-\mathbf{g}_{m,[j]}\mathbf{s}\right| ^2}{\bar{\sigma}^2_{m,j}}\bigg ).
\label{enh_pij}
\end{align}
The message passing schedule is as follows.
\begin{enumerate}
\item Initialize $\bar{p}_{ji}(\mathbf{s})=\frac{1}{|\mathbb{M}|}$ , $\forall j,i,\mathbf{s}$.
\item Compute $\bar{\mu}_{ij}$, and $\bar{\sigma}^2_{i,j}$ $\forall i,j$.
\item Compute $\bar{p}_{ji}$, $\forall j,i$. 
\end{enumerate}
Steps 2 and 3 are repeated for a fixed number of iterations. 
Damping of the messages $\bar{p}_{ji}$'s is done using a damping factor 
$\Delta \in (0,1]$ in each iteration. At the end of the iterations, 
the vector probabilities are computed as
\begin{align}
\bar{p}_{j}(\mathbf{s}) & \propto \prod_{i=1}^{Nn_r} \mbox{exp}\bigg (\frac{-\left| y_i-\bar{\mu}_{i,j}-\mathbf{g}_{i,[j]}\mathbf{s}\right| ^2}{\bar{\sigma}^2_{i,j}}\bigg ), \ j=1,2,\ldots,k.	
\end{align}
The estimates $\hat{\mathbf{w}}_j$s are obtained by choosing the signal
vector ${\bf s}\in\mathbb M$ that has the largest APPs. That is,
\begin{align}
\hat{\mathbf{w}}_j = \underset{\mathbf{s} \in \mathbb{M}}{\text{argmax}} \ \bar{p}_j(\mathbf{s}).
\end{align}
The estimate of the active antenna index in the $j$th used time slot is 
obtained from $\hat{\mathbf{w}}_j$, which is then demapped to obtain the 
antenna index bits. The non-zero entries of $\hat{\mathbf{x}}_j$ are 
demapped to get the modulation symbol bits. The indices in set $\mathcal T$ 
are demapped to get the slot index bits.

\section{Results and discussions}
\label{sec4}
In this section, we present the bit error rate (BER) performance of the 
proposed STIM detection algorithms. We compare the performance of STIM 
with that of OFDM. It is noted that both STIM and OFDM use a single 
transmit RF chain (i.e., $n_{rf}=1$ for STIM and OFDM). We consider a 
frequency-selective channel with $L=2$ and exponential power delay 
profile presented in Sec. \ref{sec2c}.
	
\begin{figure}
\centering
\includegraphics[width=3.75in,height=2.75in]{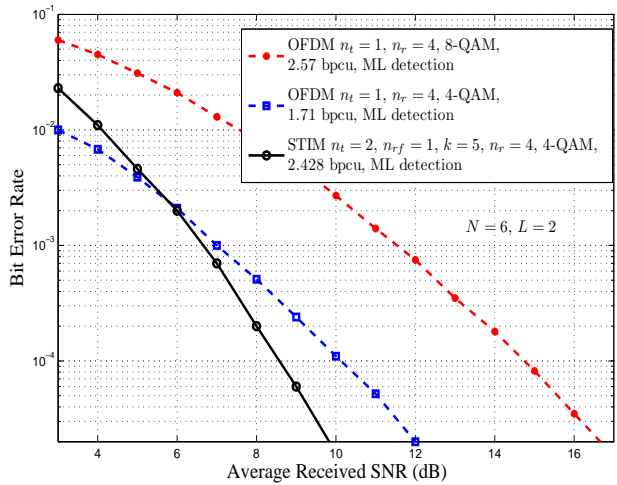}
\vspace{-6mm}
\caption{BER performance of STIM and OFDM using ML detection. STIM system
parameters are: $n_t=2$, $n_{rf}=1$, $N=6$, $k=5$, $n_r=4$, 4-QAM, and 
2.428 bpcu. OFDM system parameters are: $i)$ $n_t=1$, $n_{rf}=1$, $N=6$, 
$n_r=4$, 8-QAM and 2.57 bpcu, and $ii)$ $n_t=1$, $n_{rf}=1$, $N=6$, 
$n_r=4$, 4-QAM, and 1.71 bpcu.}
\label{Fig_ML}
\vspace{-2mm}
\end{figure}

In Fig. \ref{Fig_ML}, we present the maximum likelihood (ML) performance 
of the STIM system with $n_t=2, N=6, k=5, n_r=4,$ 4-QAM, and a
spectral efficiency of 2.428 bpcu. We compare this performance with that 
of conventional OFDM. The OFDM systems considered for comparison are:
$i)$ $n_t=1, N=6, n_r=4$, 8-QAM, and spectral efficiency of 2.57 bpcu, 
and $ii)$ $n_t=1, N=6, n_r=4$, 4-QAM, and spectral efficiency of 1.71 bpcu. 
From Fig. \ref{Fig_ML}, we observe that the STIM system outperforms the 
conventional OFDM systems. At a BER of $10^{-4}$ BER, STIM outperforms 
OFDM with 2.57 bpcu by about 6 dB, and OFDM with 1.71 bpcu by about 1.2 dB. 
This shows that STIM can achieve better performance than the conventional 
OFDM scheme as well as provide higher rate through transmit antenna and 
time slot indexing.
\begin{figure}
\centering
\includegraphics[width=3.75in,height=2.75in]{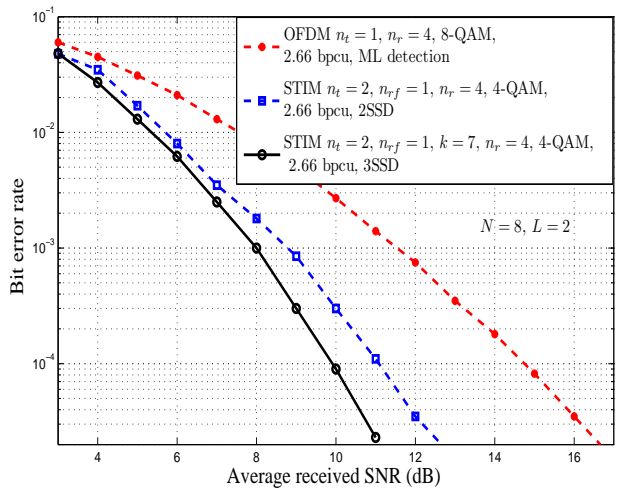}
\vspace{-6mm}
\caption{Performance comparison of STIM with 2SSD and 3SSD algorithms, 
$n_t=2$, $n_{rf}=1$, $N=8$, $k=7$, $n_r=4$, 4-QAM, and 2.66 bpcu, with 
that of OFDM with ML detection, $n_t=1$, $n_{rf}=1$, $N=8$, $n_r=4$, 
8-QAM, and 2.66 bpcu.}
\label{Fig_enhdet_N_8}
\vspace{-2mm}
\end{figure}

\begin{figure}
\centering
\includegraphics[width=3.75in,height=2.75in]{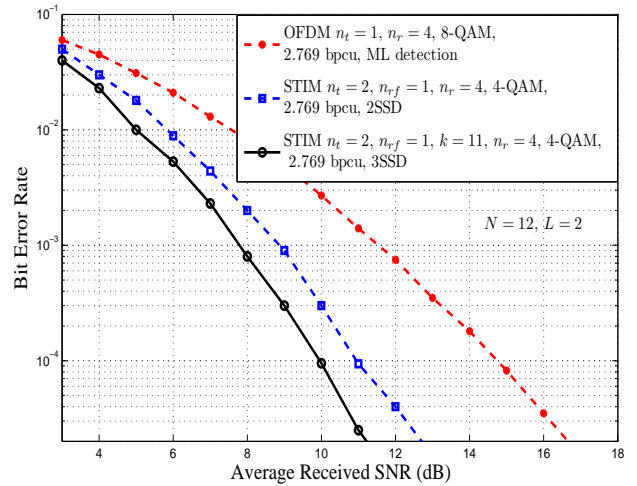}
\vspace{-6mm}
\caption{Performance comparison of STIM with 2SSD and 3SSD algorithms, 
$n_t=2$, $n_{rf}=1$, $N=12$, $k=11$, $n_r=4$, 4-QAM, and 2.769 bpcu, 
with that of OFDM with ML detection, $n_t=1$, $n_{rf}=1$, $N=12$, 
$n_r=4$, 8-QAM, and 2.769 bpcu.}
\label{Fig_enhdet_N_12}
\vspace{-2mm}
\end{figure}

In Fig. \ref{Fig_enhdet_N_8}, we present the BER performance of STIM 
detected using the proposed 2SSD and 3SSD algorithms. The value of the
damping factor used is $\Delta=0.3$. The STIM system 
considered has $n_t=2$, $N=8$, $k=7$, $n_r=4$, 4-QAM, and 2.66 bpcu 
spectral efficiency. For comparison purposes, we also present the 
performance of conventional OFDM with the same spectral efficiency 
of 2.66 bpcu, using the following system parameters: $n_t=1$, $N = 8$, 
$n_r = 4$, and 8-QAM. From Fig. \ref{Fig_enhdet_N_8}, we see that the 
performance of STIM with 2SSD algorithm is better than OFDM with ML
detection by about 3.7 dB at a BER of $10^{-4}$. We also see that STIM 
with 3SSD algorithm outperforms OFDM with ML detection by about 4.7 dB 
at a BER of $10^{-4}$. The performance of 3SSD algorithm is better than 
that of the 2SSD algorithm by about 1 dB at a BER of $10^{-4}$. This is 
because the third stage message passing in 3SSD improves the reliability 
of the detected information bits. In Fig. \ref{Fig_enhdet_N_12}, a similar 
performance gain in favor of STIM is observed for $N=12$ at 2.769 bpcu. 

\section{Conclusion}
\label{sec5}	
We proposed a new modulation scheme referred to as the space-time index 
modulation (STIM). This modulation scheme conveys information bits through 
indexing of antennas and time slots, in addition to $M$-ary modulation 
symbols. This enables STIM to achieve high spectral efficiencies. We 
showed that, for the same spectral efficiency and  single 
transmit RF chain, STIM can achieve better performance than conventional 
OFDM in frequency-selective fading channels. We also proposed two 
low-complexity detection algorithms which scale well for large dimensions. 
STIM with the proposed detection algorithms was shown to outperform 
conventional OFDM.  STIM, therefore, can be a promising modulation scheme, 
and has the potential for further investigations beyond what is reported 
in this paper. For example, generalization of indexing in 
both space and time with more than one transmit RF chain can be investigated 
as future extension to this work. Also, diversity analysis of STIM and its 
generalized schemes is another important topic for further investigations. 

\bibliographystyle{ieeetr}

\end{document}